\documentclass[11pt]{article}
\usepackage{amsfonts,amssymb,latexsym}
\usepackage{amsmath,amsgen,graphicx}
\newtheorem{claim}{Claim}{\it}{\rm}
\newtheorem{lemma}{LEMMA}[section]{\bf}{\it}
\newtheorem{prop}{PROPOSITION}{\bf}{\it}
\newtheorem{remark}{Remark}{\bf}{\rm}
\newtheorem{theorem}{THEOREM}{\bf}{\it}
\newcommand{\sphlap}{\displaystyle{\not\!\!\Delta}}
\newcommand{\sphgrad}{\displaystyle{\not\!\nabla}}
\newcommand{\ds}{\displaystyle}
\newcommand{\nab}{\nabla}
\newcommand{\half}{\frac{1}{2}}
\newcommand{\p}{\partial}
\newcommand{\A}{{\cal A}}
\newcommand{\E}{{\cal E}}
\newcommand{\Hi}{\mathbf{H}}
\newcommand{\Hc}{\mathcal{H}}
\newcommand{\M}{{\cal M}}
\newcommand{\RR}{{\mathbb R}}
\newcommand{\ZZ}{{\mathbb Z}}
\newcommand{\BS}{{\mathbb S}}
\newcommand{\mb}[1]{\mathbf{#1}}
\newcommand{\rad}{\mathrm{rad}}
\newcommand{\qed}{\par\hfill\rule{2mm}{2mm}}
\begin{document}

\title{
Scalar waves  on a\\ naked-singularity background}

\author{
John G. Stalker$^\mathbf{a}$
\and
A. Shadi Tahvildar-Zadeh\thanks{Research supported in part by the National Science Foundation grant DMS 0301207.}$^{\ ,\mathbf{b}}$
}

\date{March 2004}
\maketitle
\begin{abstract}

We obtain global space-time weighted-$L^2$ (Morawetz) and $L^4$ (Strichartz) estimates for a massless chargeless spherically symmetric scalar field propagating on a super-extremal (overcharged) 
Reissner-Nordstr\"om background. To do this we first discuss the well-posedness of the Cauchy problem for scalar fields on non-globally hyperbolic manifolds, review the role played by the Friedrichs extension, and go over the construction of the function spaces involved.  We then  show how to transform this problem to one about the wave equation on the Minkowski space with a singular potential, and  prove that the potential thus obtained satisfies the various conditions needed in order for the estimates to hold.
\end{abstract}

\section{Introduction}\label{sec:intro}

There have been many studies of the well-posedness and decay of scalar fields in a given 
space-time whose metric satisfies the Einstein equations of general relativity,
 both as a precursor to the study of the stability of that metric,
 and  as a means of probing various censorship conjectures. Although our main goal in this paper
 is to obtain estimates for scalar fields,  we need to address the issue
of well-posedness as well.

\subsection{Well-posedness of the Cauchy problem for scalar fields}

 This question arises in studying the phenomenon of singularities in general 
relativity.  The classical notion of a singularity of space-time is that of geodesic 
incompleteness. The weak cosmic 
censorship conjecture states that generically, singularities of spacetime must be hidden inside
 black holes, instead of being ``naked'', i.e. visible to distant observers.  The strong form of
 this conjecture posits that generically, spacetimes must be globally hyperbolic, i.e. possess a
 complete 
spacelike hypersurface such that every causal curve in the manifold intersects it at exactly 
one point.  Global hyperbolicity ensures that the spacetime has deterministic dynamics, since
 there is a Cauchy surface whose domain of dependence is the entire spacetime.   In the case 
where the space-time is not globally hyperbolic, which can happen for example if the metric 
is singular\footnote{The minimum regularity required of the metric for 
the local existence and uniqueness of geodesics
to hold is $C^{1,\alpha}$, $\alpha>0$, so that any point in the manifold where the metric fails to 
be $C^{1,\alpha}$
needs to be removed, which can easily result in the loss of global hyperbolicity \cite{Cla98}.  We note however, that the singularities of metrics  considered in this paper are in fact much stronger, and the curve-integrability conditions of \cite{Cla98} do not hold for these examples.} on a 
time-like curve, well-posedness of the Cauchy problem for the scalar wave 
equation has been suggested 
\cite{Wal80} as a substitute for geodesic completeness in determining how singular the 
spacetime actually is, and  perhaps to see if quantum effects have any chance of 
regularizing the dynamics and restoring predictability \cite{HorMar95}.  This is 
particularly relevant if censorship conjectures somehow fail to be true and naked 
singularities turn out to be more abundant than otherwise allowed by them.   

  Let us now  recall the set-up from \cite{Wal80}:  Let $(\M,g)$ be a 
static, stably causal space-time, i.e., one that admits a hypersurface-orthogonal Killing vector 
field $T^\mu$ whose
orbits are complete and everywhere timelike. A massless scalar field on $\M$ satisfies the 
equation 
\begin{equation}\label{eq:ur}
g^{\mu\nu}\nab_\nu \nab_\mu \psi = 0.
\end{equation}
  Suppose we specify initial data for $\psi$ on a hypersurface $\Sigma$ that is everywhere
 orthogonal to $T^\mu$.  If $\M$ is not globally hyperbolic, $\Sigma$ will not be a Cauchy
 surface and data on $\Sigma$ will determine $\psi$ only on the domain of dependence 
$D(\Sigma)$.  The aim of \cite{Wal80} was to define a physically sensible recipe for 
determining $\psi$ everywhere in $\M$.  To this end, one
rewrites
(\ref{eq:ur}) in the form
\[
\partial_t^2 \psi = \alpha D^a(\alpha D_a\psi)
\]
Here $\alpha = \sqrt{-g(T,T)}$, $t$ is the Killing parameter, and $D_a$ is the covariant
 derivative of the Riemannian metric induced on $\Sigma$.  One then views 
$$A:= -\alpha D^a(\alpha D_a)$$ as an operator on the Hilbert space 
$$\Hc = L^2(\Sigma,\alpha^{-1}d\sigma),$$ where 
$d\sigma$ denotes the induced volume form of $\Sigma$.   The point of this definition is that with respect 
to the inner product of $\Hc$,  $A$ will now be  symmetric and positive, although not 
yet self-adjoint since the initial domain of $A$ would have to consist of sufficiently 
smooth functions.  If we take this initial domain to be $C^\infty_c(\Sigma)$ 
(smooth functions of compact support on $\Sigma$) then $A$ is also densely defined, 
and the equation we want to solve is
\begin{equation}\label{eq:Apsi}
(\partial_t^2 +A) \psi = 0,\qquad \psi_{\left|_\Sigma\right.} = f,
\qquad \partial_t\psi_{\left|_\Sigma\right.} = g.
\end{equation}
It is then a classical result \cite{ReeSimII}, that the above problem is well-posed,
 provided one replaces $A$ in this
equation with (one of) its self-adjoint extension(s) $A_E$.  Such self-adjoint 
extensions are guaranteed to exist for real, symmetric operators \cite{Neu30}.  
The extension may not be unique though, which would imply that there is still an
 ambiguity about the dynamics.  This may be interpreted as having to specify
 boundary conditions for the scalar field ``on the singularity''.  One finds 
that there are three proposals for removing the ambiguity and restoring determinism 
to the dynamics of scalar fields:

{\bf (1)} One can of course restrict attention only to the cases where the 
self-adjoint extension is unique (the so-called essentially self-adjoint case), and 
declare that spacetimes where the operator $A$ is not essentially self-adjoint 
are ``quantum-mechanically singular'' \cite{HorMar95} (meaning the singularity 
remains even if quantum particle dynamics, i.e  waves, are considered in place of 
classical particle dynamics, i.e. geodesics). In this approach, the naked singularities
 present at the center in both the negative mass Schwarzschild ($m<0$) and the 
super-extremal Reissner-Nordstr\"om ($|e|>m$) spacetimes are quantum-mechanically
 singular, and nothing more can be said about the evolution of scalar fields on these 
backgrounds.

{\bf (2)} Another possibility is to characterize the singularity that leads to 
non-unique self-adjoint extensions for $A$ as having a $U(N)$ ``hair,'' \cite{IshHos99} 
where $N$ 
is the common value of the two deficiency indices $n_\pm := \dim(\ker(A^*\pm i))$ of $A$. This
 is because on the one hand there is a one-to-one correspondence between the self-adjoint extensions of $A$ 
and unitary maps from $\ker(A^*+i)$ onto $\ker(A^*-i)$ \cite{Neu30}, and
on the other hand, as
 shown in \cite{IshWal03}, any ``reasonable'' way of defining the dynamics on the whole
 spacetime, would necessarily have to arise from {\em some} self-adjoint extension of $A$. 
 We will show that the naked singularity of the super-extremal Reissner-Nordstr\"om solution
 is in this sense, quite hairy (see Remark~\ref{rem:hair}).

{\bf (3)} A third approach is to distinguish one self-adjoint extension from among all the 
possible ones, as being somehow more ``natural'' or ``physical''.  It was suggested in
 \cite{Wal80} that the Friedrichs extension of $A$, i.e. the one coming from extending
 the corresponding quadratic form, is such a natural choice. We note that the Friedrichs extension of $A$ always exists, and is unique by construction\footnote{The claimed non-uniqueness of the $H^1$-based extension (which is equivalent to the Friedrichs extension) of this operator in the case of superextremal Reissner-Nordstr\"om, \cite[\S IV.A.2]{IshHos99}, is due to the authors' error in not realizing the Sobolev space $H^1$ as the closure of $C^\infty_c$ functions under the corresponding norm (see Section 3).}. Moreover, as was shown in \cite{Seg03},  the Friedrichs extension is 
the only self-adjoint extension of $A$ under which the resulting dynamics of (\ref{eq:Apsi}) agrees  with that
 of the corresponding first-order (Hamiltonian) formulation,
\[
\partial_t \Psi = -h \Psi, 
\]
with 
\[
\Psi = \left(\begin{array}{c}\psi\\ \alpha^{-1}\partial_t \psi\end{array}\right),
\qquad h = \left(\begin{array}{cc} 0 & -\alpha \\ \alpha^{-1}A & 0\end{array}\right),
\]
obtained via the (unique) skew-adjoint extension of the corresponding operator $h$. 

In this paper we will take the third approach, and work with the
Friedrichs extension of $A$.  This is mainly because finiteness of the $H^1$
norm is used several times in the course of the proof of our estimates,
and among the extensions of $A$, the Friedrichs extension is the only
one whose domain is contained in $H^1$.  It may be that some of our
estimates hold for the other extensions of $A$ as well, but our proof
does not extend to those cases.

\subsection{Estimates for scalar fields}

The next natural question to consider for scalar waves after well-posedness is 
obtaining estimates for them.
This is relevant among other things to the question of stability of the metric as a 
solution of the 
field equations, since expanding the perturbation functions in tensor harmonics yields
 a sequence of scalar wave equations \cite{Mon74}, and having good estimates for each
 one of them seems necessary--although  not sufficient--for proving linear
 stability of the metric under a small perturbation of its data.  Moreover, the question 
of stability has an obvious connection to cosmic censorship if the solution 
to be perturbed happens to have a naked singularity.

Most of the results obtained so far in the direction of estimates
concern either boundedness of the field \cite{KayWal87,Whi89,
Bey01} or its dispersion, i.e. the decay rate, with respect to the foliation parameter, of the 
supremum of the field on the slices of a given time-like or  null foliation of
a portion of the space-time 
 \cite{Pri72,Bic78,KoyTom01a,KoyTom01b,FKSY02,CYDL03,Poi03,MacSta04,DafRod03}.
  In this paper we take a different approach to this problem, by
 obtaining estimates that bound the {\em global} space-time $L^4$-norm  (and weighted-$L^2$ 
norm) of the field, in terms 
of appropriate norms of its data on a given Cauchy hypersurface (see \cite{BluSof03,BluSof03b} for
a similar approach).  Such Strichartz (resp. Morawetz) estimates,
 as they are known in the literature,
 are a staple of the modern  theory of partial differential equations, and have  proved
 very useful in the analysis of nonlinear evolution problems.  We will be looking at
 the simplest non-trivial situation possible, namely, a  spherically symmetric scalar field 
on a given spherically symmetric and static background space-time\footnote{It is possible to 
extend our result to fields that are a finite sum of spherical harmonics, as in \cite{FKSY02}, but we
do not pursue it here since this approach is not likely to yield an estimate for the general,
 nonsymmetric case.}.
Furthermore, the background here is allowed 
to have a naked singularity, the main example we have in mind being the super-extremal
 (naked) Reissner-Nordstr\"om solution of the Einstein-Maxwell system.  Our main result
 is
\begin{theorem}\label{thm:intro}
Let $(\M,g)$ be the Reissner-Nordstr\"om space-time manifold, with mass $m$ and charge $e$, 
such that
\[
|e|> 2m,
\]
and let $\psi$ be a  massless, chargeless, spherically symmetric,
 scalar field on $\M$.  Then there exists a constant $C>0$ 
(which may
depend on $e$ and $m$) such that 
\begin{equation}\label{est:intro}
\| \psi\|_{L^4(d\mu_g)} + \|\rho^{-1}\psi\|_{L^2(d\mu_g)} \leq C
\mb{E}_{1/2}(\psi),
\end{equation}
where $\rho$ denotes the area-radius coordinate on $\M$, defined in Section~2 and
 $\mb{E}_{1/2}(\psi)$ is the 
conserved $\half$-energy of the field, defined in Section~\ref{sec:energy}.
\end{theorem}
The outline of this paper is as follows:  In Section 2 we introduce the family of static
 spherically-symmetric Lorentzian manifolds, define a global system of coordinates on them,
 and  write down the evolution equation satisfied by a scalar field on such a manifold.  
In Section 3 we introduce the notion of self-adjoint extension that is necessary to make 
that evolution problem well-posed, and define the function spaces that are going to be
 used.  In Section 4 we state the various conditions that need to be satisfied by the 
metric coefficients of a manifold in the family we are considering, such that a scalar 
field on it will satisfy the  estimate (\ref{est:intro}).  This is then proved by 
transforming the problem to one about the flat wave equation with a potential,  
appealing to 
the result in \cite{BPST2}, and transforming back to the original problem.  The 
last section contains the proof of the fact that the metric coefficients of the 
super-extremal Reissner-Nordstr\"om 
solution satisfy the above mentioned conditions.  This is accomplished by reformulating
 the problem in the language of real algebraic curves and appealing to the compactness result
 of~\cite{Sta03b}.
\section{Scalar fields on static spherically symmetric\\ backgrounds}\label{sec:scal}
\subsection{The space-time}
Consider a four-dimensional connected spherically symmetric static space-time
$(\mathcal{M},g)$.  More precisely on $\M$ we assume a time-like action of $\RR$
and a space-like action of $SO(3,\RR)$ commuting with it.  These
actions should be without fixed points, except that at most one $\RR$-orbit
is allowed to be $SO(3,\RR)$-fixed, in which case it will be
called the time axis.  We restrict our attention to the cases where  the $SO(3,\RR)$-orbits
off the time axis are spheres.

\subsection{The coordinate system}
Let $T=T(p)$ denote the Killing vector
generating the $\RR$-action at $p\in\M$ and let $A=A(p)$ denote the area of the
$SO(3,\RR)$-orbit passing through the point $p\in\M$. We define two $\RR\times SO(3,\RR)$-invariant 
functions on $\M$ as follows: 
\[
\alpha := \sqrt{-g(T,T)},\qquad
\rho := \sqrt{A/4\pi}.
\]
The quotient space $(\mathcal{Q},\overline{g}):=(\mathcal{M},g)/SO(3,\RR)$
is a two-dimensional Lorentzian manifold.\footnote{Possibly a
manifold with boundary if there is a time axis.}
We can assign coordinates $t,r$ on $\mathcal{Q}$ in such a way that
\begin{displaymath}
\overline{g}_{ab}\,dy^a\,dy^b = \alpha^2(r)\left(-dt^2+dr^2\right).
\end{displaymath}
The Killing field is clearly just $\partial_t$
in these coordinates.  We pull back
$t$ and $r$ to  $\mathcal{M}$.  It is easy to see that
we can assign angular coordinates $\Omega = (\theta,\phi)$ on $\mathcal{M}$ in
such a way that the metric becomes
\begin{equation}\label{metricform}
g_{\mu\nu}\,dx^\mu\,dx^\nu = \alpha^2(r) \left(-dt^2+dr^2\right)
+ \rho(r)^2 (d\phi^2+\sin^2\phi d\theta^2),
\end{equation}
and the action of $SO(3,\RR)$ is the usual action on the unit
sphere.   These coordinates are unique up to translations in the
$t,r$ coordinates and rotations in the $\theta,\phi$ coordinates.
If there is a time axis we can then arrange that $r=0$ there.

We note that in these coordinates the volume form of the spacetime is
\[
d\mu_g = \alpha^2 \rho^2  d\Omega dr dt,
\]
while the induced volume form on the Cauchy hypersurfaces $t = \mbox{const.}$ is
\[
d\sigma = \alpha \rho^2  d\Omega dr.
\]
Here $$d\Omega = \sin\phi d\theta d\phi$$ denotes the volume form on the standard 2-sphere. We also note that there are at least two other volume forms on these hypersurfaces that are natural to 
consider, namely
\[
d\sigma' = \alpha d\sigma = \alpha^2 \rho^2  d\Omega dr\qquad\mbox{ and }
\qquad d\sigma'' = \alpha^{-1} d\sigma = \rho^2  d\Omega dr.
\]
The significance of $d\sigma'$ is that $L^p$-norms based on it behave as one would
expect with respect to the $t$-foliation:  For $f:\M \to \RR$
and all $p\geq 1$,
\[
\| f \|_{L^p(d\mu_g)} = \|\ \| f \|_{L^p(d\sigma')} \|_{L^p(dt)}.
\]
The significance of $d\sigma''$ will become clear in the next section.

\subsection{The wave equation}
Consider the action functional $\A = \half \int_\M L d\mu_g$ corresponding to a real massless scalar field $\psi:\M\to\RR$.  The Lagrangian density is 
$$L =  g^{\mu\nu}\nab_\mu \psi \nab_\nu \psi.$$
  In the above-given coordinates
\[
L = -\alpha^{-2} \psi_t^2 + \alpha^{-2} \psi_r^2 + \rho^{-2}|\sphgrad
\psi|^2,
\]
where $\sphgrad$ denotes the unit-spherical gradient, and 
\[
|\sphgrad \psi|^2 = |\partial_\phi \psi|^2 + \frac{1}{\sin^2\phi} |\partial_\theta \phi|^2.
\]
  A scalar field $\psi$ that is a stationary  point of the  action $\A$, subject to 
a given set of initial values $(\psi_0,\psi_1)$ on the Cauchy hypersurface $t=0$, satisfies the 
following Cauchy problem
\begin{equation}\label{eq:psi}
\partial_t^2 \psi + A \psi = 0,\qquad \psi(0) = \psi_0,\qquad
\partial_t\psi(0) = \psi_1,
\end{equation}
where 
\begin{equation}\label{def:A}
A := - \frac{1}{\rho^2}  \partial_r (\rho^2 \partial_r) -
\frac{\alpha^2}{\rho^2} \sphlap,
\end{equation}
with $\sphlap$ denoting the Laplace-Beltrami operator on the unit 2-sphere. Note that the operator $A$ is symmetric
and positive definite with respect to the inner product given by the volume form $d\sigma''$, 
i.e. for $\phi,\psi \in C^\infty_c(\Sigma_t)$,
\[
\int_{\Sigma_t} \phi A \psi\ d\sigma'' = \int_{\Sigma_t}\psi A\phi\  d\sigma'',
\]
and
\[
\int_{\Sigma_t} \phi A \phi\ d\sigma'' = 
\int_{\Sigma_t} \phi_r^2 + \frac{\alpha^2}{\rho^2} |\sphgrad \phi|^2\
d\sigma'' \geq 0,
\]
with equality only if $\phi\equiv 0$.  

\section{Self-Adjoint Extensions, Energy, and Sobolev Norms}\label{sec:energy}
As explained in Section~\ref{sec:intro}, the evolution
problem (\ref{eq:psi}) is only meaningful if the operator $A$ there is replaced
by (one of) its self-adjoint extension(s) $A_E$. In the event that the self-adjoint 
extension is not unique, the 
one we choose to pick is the Friedrichs extension $A_F$, obtained by extending the 
 quadratic
form naturally associated with the operator $A$, namely
\begin{equation}\label{def:Q}
Q_A(\phi) := \int_{\Sigma_t} \phi_r^2 + \frac{\alpha^2}{\rho^2}
|\sphgrad \phi|^2\  d\sigma''.
\end{equation}
Thus the Cauchy problem we are actually studying is the following
\begin{equation}\label{eq:psiFri}
\partial_t^2 \psi + A_F \psi = 0,\qquad \psi_{|_\Sigma} = f,\qquad \partial_t \psi_{|_\Sigma} = g.
\end{equation}

Let $T_{\mu\nu}$ be the energy tensor of the field, defined as
\[
T_{\mu\nu} = \nab_\mu \psi \nab_\nu \psi - \half g_{\mu\nu}L.
\]
In particular, in the above given coordinates
\[
T_{00} = \half \psi_t^2 + \half \psi_r^2 + \frac{\alpha^2}{2\rho^2}
|\sphgrad \psi|^2.
\]
Let $X$ be a time-like Killing vector field for $\M$.  Then the one-form 
$P$ defined by $P(Y) := T(X,Y)$ is divergence-free, i.e. ${}*d*P = 0$ and thus $*P$ is a 
conserved current.  
In particular, let
 $X= \partial_t$, then $P_0 = T_{00}$ and $P_i = T_{0i}$, so that the only nonzero 
component of $*P$ is
$(*P)_{123} = \rho^2 T_{00}$, and the conserved quantity is the {\em energy}
\[
\mathbf{E}[\psi]:=\int_{\Sigma_{t}} \rho^2 T_{00}\ dr d\Omega = 
\int_{\Sigma_{t}} T_{00}\  d\sigma''.
\]
Thus  using the $d\sigma''$ volume form, $T_{00}$ is identified with the 
{\em energy density} of the field.  

It therefore makes sense to also use $d\sigma''$ to define Sobolev spaces on the $\Sigma_t$'s:  
Let $\Hi^1(\Sigma_t)$ denote the 
completion of smooth compactly supported functions on $\Sigma_t$ with respect to the norm
\[
\| f\|_{\Hi^1}^2 := \int_{\Sigma_t} |f_r|^2 + \frac{\alpha^2}{\rho^2}
|\sphgrad f|^2 \ d\sigma''.
\]
We define $\Hi^0(\Sigma_t)$ in the same way, i.e. completion with respect to the norm
\[
\| f \|_{\Hi^0}^2 := \int_{\Sigma_t} |f|^2\ d\sigma''.
\]
The Sobolev spaces $\Hi^s(\Sigma_t)$ for $0<s<1$ are then defined via interpolation between
 the above
two spaces, and one uses duality to define them also for $-1\le s<0$.   We note that by these 
definitions,
\[
Q_A(\phi) = \| \phi\|_{\Hi^1(\Sigma_t)}^2,
\]
and moreover
\[
\mathbf{E}[\psi] = \half \left(\| \psi \|_{\Hi^1(\Sigma_t)}^2 + 
\| \psi_t \|_{\Hi^0(\Sigma_t)}^2\right).
\]
More generally, for $0 \le s \le 1$ we can define the $s$-energies
\[
\mathbf{E}_s[\psi] := \half\left( \|\psi\|_{\Hi^s(\Sigma_t)}^2 + 
\| \psi_t \|_{\Hi^{s-1}(\Sigma_t)}^2 \right),
\]
and it is not hard to see that they are all conserved under the flow (\ref{eq:psiFri}): 
\[
\frac{d}{dt} \mathbf{E}_s[\psi] = 0.
\]

\section{Transferring to Minkowski Space}
It is clear that if we can  view (\ref{eq:psi}) as an evolution in Minkowski space, we may then be able
 to apply known theorems in order to get the  estimates we want. This is of course only possible if the 
manifold $\M$ has at least the same topology as $\RR^4$ (or $\RR^4$ minus a line,
in order to allow a singular time axis). We will be making this assumption from now on, and denote 
the coordinates on $\RR^4$ by the same letters as those on $\M$, namely
 $(t,r,\Omega)\in \RR\times \RR^+\times \BS^2$, and a $t$-slice in $\RR^4$ is still denoted by $\Sigma_t$. 
Let us define $u:\RR^4\to \RR$ by
\[
u(t,r,\Omega):= \frac{\rho(r)}{r} \psi(t,r,\Omega).
\]
Note that
\begin{equation}\label{equiL2}
\| u \|_{L^2(\Sigma_t)}^2 = \int_{\Sigma_t} |u|^2 r^2 dr d\Omega = 
\int_{\Sigma_t} |\psi|^2 d\sigma'' = \|\psi\|_{\Hi^0(\Sigma_t)}^2.
\end{equation}

We can easily check that for $\psi\in C^\infty_c(\Sigma_t)$ we can perform an integration by parts
to obtain
\begin{eqnarray}
Q_A(\psi) &=& \int_{\Sigma_t} \left[ u_r^2 + \frac{\alpha^2}{\rho^2} |\sphgrad u|^2 + 
\frac{\rho^2}{r^2}[\partial_r(\frac{r}{\rho})]^2 u^2 +
 \frac{\rho}{r}\partial_r(\frac{r}{\rho})\partial_r (u^2) \right] r^2 dr\,d\Omega\nonumber\\
& = & \int_{\Sigma_t} \left[ u_r^2 + \frac{\alpha^2}{\rho^2} |\sphgrad u|^2 + 
V(r) u^2 \right] r^2 dr d\Omega =:  Q_B(u),\label{equiQ}
\end{eqnarray}
where the ``potential" $V$ is defined by 
\begin{equation}\label{def:V}
V(r) = \rho''(r)/\rho(r).
\end{equation}
Let $B$ denote the operator whose associated quadratic form is $Q_B$ defined above, i.e.
\[
B := -\frac{1}{r^2}\partial_r(r^2\partial_r) - \frac{\alpha^2}{\rho^2}
\sphlap + V.
\]
We then have that 
\[
\partial_t^2 + A = \frac{r}{\rho} (\partial_t^2 + B) \frac{\rho}{r}.
\]
Note that the first term in the definition of $B$ coincides with the radial flat Laplacian. 
In particular, if $\psi$ satisfying (\ref{eq:psi}) is spherically symmetric, i.e. $\psi = \psi(t,r)$, then
 $u$ can be viewed as a {\em radial} solution of the flat wave equation with a potential
\begin{equation}\label{eq:u}
u_{tt} - u_{rr} - \frac{2}{r} u_r + V(r) u  = 0,\qquad u(0,r) =
f(r),\qquad u_t(0,r) = g(r),
\end{equation}
with $f = \rho \psi_0/r $ and $g = \rho \psi_1/r$.  

Since by (\ref{equiQ}) the quadratic forms corresponding to $A$ and $B$ are equivalent,
we can  identify the Friedrichs extensions $A_F$ and $B_F$ 
of  $A$ and $B$, as well 
as the Sobolev spaces based on them, i.e. let $\Hc^s$ denote
the $B$-based Sobolev space, defined to be the completion of smooth compactly supported functions 
on $\RR^3\setminus\{0\}$ with respect to the norm
\[
\|f\|_{\Hc^s} := \| (B_F)^{s/2} f \|_{L^2}.
\]
We then have that for $|s|\leq 1$
\[
\| u \|_{\Hc^s} = \| \psi \|_{\Hi^s}.
\]
This is because the $L^2$ spaces agree by (\ref{equiL2}) while the $H^1$ spaces agree because of (\ref{equiQ}). 
On the other hand, if we let 
\[
P := -\Delta+V = - \frac{1}{r^2}\partial_r(r^2\partial_r) - \frac{1}{r^2} \sphlap +V,
\]
 then it is clear that
on the subspace of radial functions, $B_F$ coincides with $P_F$ (the Friedrichs extension of $P$), and so do the 
corresponding Sobolev spaces, which we will denote by $\Hc_\rad$.  In particular if we define the $s$-energy of $u$ to be
\[
\E_s[u] := \| u(t) \|_{\Hc^s_\rad} + \|u_t(t)\|_{\Hc^{s-1}_\rad},
\]
then 
\begin{equation}\label{Eequiv}
\E_s[u] = \mathbf{E}_s[\psi]
\end{equation}
 and it is conserved by the flow (\ref{eq:u}).

In \cite{BPST2} it was shown that solutions to 
\begin{equation}\label{eq:P}
(\partial_t^2 + P_F)u = 0
\end{equation}
satisfy certain
 weighted-$L^2$ (Morawetz) and $L^p$ (Strichartz)
spacetime estimates given below, provided the potential $V$ meets certain criteria. 
 In the case of a {\em radial} potential $V =V(r)$ on $\RR^3$ these criteria reduce to the following three conditions
 on $V$:
\begin{eqnarray}\label{cond1}
\sup_{r>0} r^2 V(r) & < & \infty,\\
\inf_{r>0} r^2 V(r)& > & -1/4, \label{cond2}\\
\sup_{r>0} r^2 \frac{d}{dr}(r V(r)) &<& 1/4. \label{cond3}
\end{eqnarray}
The version of the main result in \cite{BPST2} for radial potentials (and general data) is as follows:
\begin{theorem}\label{thm:bpst2}
Let $V\in C^1((0,\infty),\RR)$ satisfy (\ref{cond1},\ref{cond2},\ref{cond3}), and let $P := -\Delta + V(|x|)$
where $\Delta$ is the Laplace operator with domain $C^\infty_c(\RR^3\setminus\{0\})$, and let $P_F$ denote its Friedrichs extension. 
Then
there exists a constant $C$, depending only on the quantities on the
left in (\ref{cond1},\ref{cond2},\ref{cond3}), such that any
solution $u$ of (\ref{eq:P}) satisfies
\begin{equation}\label{est:morstr}
\| r^{-1} u \|_{L^2(\RR^4)} + \|  u\|_{L^4(\RR^{4})} \leq C \E_{1/2}[u].
\end{equation}
\end{theorem}

Now, the space-time $L^p$ norms of $\psi$ and $u$ are related in the following way:
\[
\|\psi\|_{L^p(\M)}^p = \int_\M |\psi|^p d\mu_g =  
\int_{-\infty}^\infty\int_{\BS^2}\int_0^\infty \alpha^2 
\left(\frac{r}{\rho}\right)^{p-2}\ |u|^p \ r^2 dr\, d\Omega\, dt. 
\]
In particular 
\begin{equation}\label{L4equiv}
\|\psi\|_{L^4(\M)}^4 = \int_{\RR^4} \left(\frac{\alpha r}{\rho}\right)^2
|u|^4 d^3x dt.
\end{equation}
Similarly
\begin{equation}\label{L2equiv}
\| \frac{\psi}{\rho}\|_{L^2(\M)}^2 = \int_{\RR^4} \left(\frac{\alpha r}{\rho}\right)^2 
\frac{|u|^2}{r^2} 
d^3x dt.
\end{equation}

We thus have proved the following theorem regarding estimates for the scalar field
 $\psi$ on $\M$:

\begin{theorem}\label{thm:main}
Let $\M$ be a Lorentzian manifold that is homeomorphic to~$\RR^4$, admitting a timelike $\RR$ 
action and a spacelike $SO(3,\RR)$ action commuting with it, in such a way that exactly one 
$\RR$-orbit, called $\Gamma$, is $SO(3,\RR)$-fixed.  Let 
$(t,r,\Omega)\in \RR\times\RR^+\times \BS^2$ be the coordinate system on $\M$ as in 
Section~\ref{sec:scal}, with $\Gamma = \{r=0\}$.  Let $g$ be a Lorentzian metric 
on $\M$ that is of 
the form (\ref{metricform}), is $C^3$ outside $\Gamma$, 
 and is such that the functions $\rho$ and $\alpha$ satisfy 
the following conditions
\begin{itemize}
\item[\rm(i)] $\sup_{r>0} (r^2V) < \infty$
\item[\rm(ii)] $\inf_{r>0} (r^2 V) > -1/4$
\item[\rm(iii)] $\sup_{r>0} (r^2\frac{d}{dr}(rV)) < 1/4$
\item[\rm(iv)] $\inf_{r>0} (\ds\frac{\rho}{\alpha r}) > 0$
\end{itemize}
where 
\[
V(r) := \frac{1}{\rho}\frac{d^2\rho}{dr^2}.
\]
Then there exists a constant $C>0$, depending only on the quantities on the
left in the conditions above,
such that any spherically symmetric solution of \[ \partial_t^2 \psi + A_F \psi = 0 \] satisfies
\begin{equation}\label{est:main}
\| \rho^{-1} \psi\|_{L^2(\M)} + \| \psi\|_{L^4(\M)} \leq C \mathbf{E}_{1/2}[\psi].
\end{equation}
\end{theorem}

{\em Proof:}  We simply observe that by (\ref{L4equiv},\ref{L2equiv}),
\[
\| \rho^{-1}\psi\|_{L^2} + \| \psi \|_{L^4} \leq \frac{1}{d} \|r^{-1} u \|_{L^2} + \frac{1}{\sqrt{d}} \| u \|_{L^4}
\] 
where $d$ denotes the quantity on the left in condition (iv).  The result then 
follows from (\ref{est:morstr}) and (\ref{Eequiv}).
\qed

\section{Super-extremal Reissner-Nordstr\"om}

The Reissner-Nordstr\"om manifold $(\M,g)$ is the static, spherically 
symmetric solution of Einstein-Maxwell equations.  It is characterized
 by two parameters:  mass $m$ and charge $e$.
It can be shown that for a metric of the form (\ref{metricform}) to satisfy the Einstein-Maxwell system, one must have
\begin{equation}\label{ode}
\frac{d\rho}{dr} = \alpha^2
\end{equation}
and
\begin{equation}\label{def:alpha}
\alpha = \sqrt{1 - \frac{2m}{\rho} + \frac{e^2}{\rho^2}}.
\end{equation}
It thus follows that when $|e|<m$ the function $\rho$ is bounded away from zero and there is no time axis.  This is called the sub-extremal (black hole) case.
  When $|e|>m$ the manifold $\M$ has the same topology as $\RR^4$ minus a line,
 the time axis is at $r=0$ and the metric $g$ is highly singular there.  This is the super-extremal (naked) case of Reissner-Nordstr\"om.

The ODE (\ref{ode}) can be solved to express the ``tortoise" coordinate $r$ as a function of $\rho$.  In the super-extremal case,
\begin{equation}\label{longugly}
r(\rho) = r_0 + \rho + m\log(\frac{\rho^2-2m\rho+e^2}{e^2-m^2}) +
 \frac{2m^2-e^2}{\sqrt{e^2-m^2}}\tan^{-1}\frac{\rho-m}{\sqrt{e^2-m^2}}.
\end{equation}
We choose the constant $r_0$ such that  $r(0) =0$. Since $r$ is an increasing function of $\rho$ this 
implicitly defines $\rho$ as a function of $r$.
It is also easy to compute from the ODE (\ref{ode}) that
\[
\lim_{r\to 0} \frac{\rho}{r^{1/3}} = (3e^2)^{1/3},\qquad \lim_{r\to \infty} \frac{\rho}{r} = 1.
\]
and as a result
\[
\lim_{r\to 0} r^{1/3} \alpha = (e/3)^{1/3},\qquad \lim_{r\to \infty} \alpha = 1,
\]
so that condition (iv) of Theorem~\ref{thm:main} is clearly satisfied since
\[ \frac{\rho}{\alpha r} \to \infty\mbox{ as }r\to 0,\qquad \frac{\rho}{\alpha r} \to 
1 \mbox{ as } r \to \infty.
\]

To prove Theorem~\ref{thm:intro}, it thus remains to check that the function
\[
	V = \frac{1}{\rho}\frac{d^2\rho}{dr^2} =
	\frac{2m}{\rho^3} - \frac{2e^2 + 4m^2}{\rho^4}
	+ \frac{6me^2}{\rho^5} - \frac{2e^4}{\rho^6},
\]
satisfies conditions (i-iii) of Theorem~\ref{thm:main}. In the case of condition (iii) this does not appear to be 
an easy task,  because of the
 transcendental relation (\ref{longugly}) between $r$ and $\rho$, and the dependence on the two additional variables
$e$ and $m$.  To overcome these difficulties, first we exploit the inherent scaling in the problem to 
eliminate $e$ (effectively setting it equal to one)
and  then by forgetting the relationship between  $r$ and $\rho$ and treating them as independent quantities,
 we transform the problem into one about
polynomials in three variables, which we then solve by utilizing a few basic tools from the theory of
 real algebraic curves. 

\begin{remark}\rm The lower bound on the charge-to-mass ratio in  Theorem~\ref{thm:intro} is not sharp.  The actual
ratio for which condition (iii) of Theorem~\ref{thm:main} starts to be violated is  less than 2  (It is 
not hard to see  that this condition is  violated for the ratios close to 1).  We have not 
attempted to find the precise cut-off point, since it is not known whether condition (iii) is necessary 
for the estimate (\ref{est:main}) to hold. We also note that
 the charge-to-mass ratios that occur in nature, such as in 
elementary particles, are in fact very large\footnote{The charge $e$ and mass $m$ appearing in
 (\ref{def:alpha}) are 
in geometric rationalized units, with $G = c = \hbar = 1$. The corresponding quantities in unrationalized 
electrostatic units $e^*$ and $m^*$ are related to $e$ and $m$ in the following way:  $m = \frac{G}{c^2} m^*$ and
$e = \frac{\sqrt{G}}{c^2} e^*$.  It follows that $e/m \approx 10^{21}$ for an electron and $\approx 2\times 10^{18}$ for a proton. }.
\end{remark}

We perform the scaling first:  Let
\[
x := \frac{r}{|e|},\qquad y := \frac{\rho}{|e|},\qquad z := \frac{m}{|e|},\qquad v := e^2 V
\]
We then have
\[
\alpha^2 = 1 - \frac{2z}{y} + \frac{1}{y^2}.
\]
Let $y = \phi_z(x)$ denote the solution to the following initial value problem
\[
\frac{dy}{dx} = \alpha^2(y,z),\quad y(0) = 0.
\]
The inverse function of $\phi_z$ is in fact explicit:
\[
\phi_z^{-1}(y) =  y + \log(y^2 - 2 yz +1) +
 \frac{2z^2-1}{\sqrt{1-z^2}}\tan^{-1}\frac{y\sqrt{1-z^2}}{1-yz}.
\]
Note that since we are considering the super-extremal case, the parameter~$z$ ranges from 0 to 1.

From the ODE it is clear that 
\begin{equation}\label{phiasymp}
\lim_{x\to 0} \frac{(\phi_z(x))^3}{x} = 3,\qquad \lim_{x\to \infty}\frac{\phi_z(x)}{x} = 1.
\end{equation}
We have
\[
v(y,z) = \frac{1}{y} \frac{d^2 y}{dx^2} =  \frac{2z}{y^3}-\frac{2+4z^2}{y^4} + \frac{6z}{y^5}-\frac{2}{y^6} = 
\frac{2}{y^6}(yz-1)(y^2-2yz+1).
\]  
From (\ref{phiasymp}) it follows that
\begin{eqnarray}
\lim_{x\to 0}x^2 v(\phi_z(x),z) & = & -\frac{2}{9},\label{vasmyp0}\\
\lim_{x\to\infty} x^2 v(\phi_z(x),z) & =  & 0,\label{vasympinf}
\end{eqnarray}
and thus conditions (i) and (ii) of Theorem~\ref{thm:main} are satisfied since 
$-\frac{2}{9}>-\frac{1}{4}$ and $v$ is clearly 
negative and increasing for $y<1/z$, positive for $y>1/z$, and smooth away from $x=0$.

\begin{remark}\label{rem:hair}\rm
Note however that since $-\frac{2}{9}<\frac{3}{4}$,  the operator
 $$Q:= -\p_x^2 -\frac{2}{x}\p_x + v(\phi_z(x),z)$$ with domain $C^\infty_c(\RR^+)$ will have non-unique self-adjoint extensions, i.e. is 
limit-circle at zero (see \cite[Appendix to X.1]{ReeSimII}).  In this sense, with respect to {\em spherically symmetric} scalar waves, the naked singularity at $x=0$ of the super-extremal Reissner-Nordstr\"om 
solution has a $U(1)$ hair.  Note also that the angular momentum contribution to the full Laplace-Beltrami
operator of this manifold is
\[
\frac{\alpha^2(\phi_z(x),z)}{(\phi_z(x))^2}\ell(\ell+1)
\]
where $\ell$ is the spherical harmonic degree.  Since this behaves like $x^{-4/3}$ near
 the origin,  its
addition to $Q$ does not change anything as far as self-adjoint extensions are concerned. 
 It thus follows that with respect to scalar waves in general,  this naked singularity is
 ``infinitely hairy," i.e. the corresponding unitary group is certainly infinite-dimensional\footnote{This provides  us with another reason for choosing the Friedrichs extension instead.}.
\end{remark}
Consider now the expression in condition (iii) of Theorem~\ref{thm:main}:
\begin{eqnarray*}
x^2\frac{d}{dx}(xv(\phi_z(x),z)) &=&
 x^2v(\phi_z(x),z)+ x^3 \frac{\partial v}{\partial y}(\phi_z(x),z)
\alpha^2(\phi_z(x),z)\\
& = & w(x,\phi_z(x),z),
\end{eqnarray*}
where                                                                                                                                                                                                                                        
\[
	\begin{array}{rcl}
	w(x,y,z) &:=& 	\left(-\frac{6z}{y^4} + \frac{8 + 28z^2}{y^5}
	- \frac{52z + 32z^3}{y^6} + \frac{20 + 76z^2}{y^7}
	- \frac{54z}{y^8} + \frac{12}{y^9}\right)x^3
	\cr && {}
	+ \left(\frac{2z}{y^3} - \frac{2 + 4 z^2}{y^4}
	+ \frac{6z}{y^5} - \frac{2}{y^6}\right)x^2.
	\end{array}
\]
Let us define the function $j_z:\RR^+\to\RR^+$ by
\begin{equation}\label{def:jz}
j_z(y) = w(\phi_z^{-1}(y),y,z).
\end{equation}
A long, routine computation of Taylor series shows that near $y=0$,
\begin{equation}\label{taylor}
j_z(y) = \frac{2}{9} - \frac{4-z^2}{270} y^2 + O(y^3),
\end{equation}
\begin{equation}\label{taylor2}
j'_z(y) = - \frac{4-z^2}{135} y + O(y^2),
\end{equation}
and
\begin{equation}\label{taylor3}
j''_z(y) = - \frac{4-z^2}{135} + O(y).
\end{equation}
while, for large $y$,
\begin{equation}\label{taylor4}
j_z(y) = -4zy^{-3} - 28zy^{-4}\log y + O(y^{-4})
\end{equation}
and
\begin{equation}\label{taylor5}
j'_{z}(y) = 12zy^{-4}  + 112zy^{-5}\log y + O(y^{-5}).
\end{equation}
The implied constants in $O$-terms are locally uniform in~$z$.
Thus $j_z$ has a local maximum at $y=0$, for $0\leq z<1$.
We wish to prove that, for $z \leq 1/2$, this is in fact the global maximum of $j_z$.  Since it is easy to see
that $j_z\to 0$ as
$y\to\infty$, it suffices to show that the only local maximum is the
one at $y=0$.  To this end we first show that 
\begin{prop}\label{prop:nospi}
$j_z$  has no stationary
points of inflection as long as $z\leq 1/2$.
\end{prop}

{\em Proof:} We compute
\[
j'_z(y) = \frac{\partial w}{\partial y}(\phi_z^{-1}(y),y,z) + \frac{\partial w}{\partial x}(\phi_z^{-1}(y),y,z)\frac{1}{\alpha^2(y,z)} =  q_1(\phi_z^{-1}(y),y,z),
\]
where
\[
\begin{array}{rcl}
q_1(x,y,z) & := &  \left( {\frac {24 z}{{y}^{5}}}-{\frac {40+140\,{z}^{2}}{{y}^{6}}}+{
\frac {312\,z+192\,{z}^{3}}{{y}^{7}}}\right.\\
&&\quad-\left. {\frac {140+532\,{z}^{2}}{{y}^{
8}}}+{\frac {432 z}{{y}^{9}}}-\frac{108}{y^{10}} \right) {x}^{3}\\
&&{} + \left( 
-{\frac {24z}{{y}^{4}}}+{\frac {64\,{z}^{2}+32}{{y}^{5}}}-\,{
\frac {120 z}{{y}^{6}}}+\frac{48}{y^{7}} \right) {x}^{2}\\
&&{} + \left( {\frac {4z}
{{y}^{3}}}-\frac{4}{{y}^{4}} \right) x,
\end{array}
\]
and similarly we compute
\[
j_z''(y) = \frac{1}{\alpha^2(y,z)} q_2(\phi_z^{-1}(y),y,z),
\]
where
\[
\begin{array}{rcl}
q_2(x,y,z) & := & 
	\left( -{\frac {120 z}{{y}^{4}}}+{\frac {240+1080\,{z}^{2}}{{y}^{5}}}
+{\frac {-3024\,{z}^{3}-2784\,z}{{y}^{6}}}\right.\\
&&\quad\left. +{\frac {1360+9464\,{z}^{2}
+2688\,{z}^{4}}{{y}^{7}}}+{\frac {-8312\,z-9856\,{z}^{3}}{{y}^{8}}}\right.\\
&&\quad+\left. {
\frac {2200+12032\,{z}^{2}}{{y}^{9}}}-\,{\frac {6048z}{{y}^{10}}}+\frac{1080}
{y^{11}} \right) {x}^{3}\\
&&{} + \left( {\frac {168 z}{{y}^{3}}}+{\frac {
-932\,{z}^{2}-280}{{y}^{4}}}+{\frac {2072\,z+1216\,{z}^{3}}{{y}^{5}}}\right.\\
&&\quad\left. +
{\frac {-3356\,{z}^{2}-916}{{y}^{6}}}+\,{\frac {2688z}{{y}^{7}}}-\frac{660}
{y^{8}} \right) {x}^{2}\\
&&{} + \left( -{\frac {60 z}{{y}^{2}}}+{\frac {80+
152\,{z}^{2}}{{y}^{3}}}-\,{\frac {284z}{{y}^{4}}}+\frac{112}{y^{5}}
 \right) x\\
&&{}+{\frac {4z}{y}}-\frac{4}{y^{2}}.
\end{array}
\]
To prove the statement of the proposition, it is enough to show that the zero-sets of
 $q_1(x,y,z)$ and $q_2(x,y,z)$ are disjoint. The strategy is to forget about the 
transcendental relationship between $x$ and $y$ and treat them as independent variables, 
 in order to be able to use results from the theory of real algebraic curves.

At a stationary point of inflection both $j'_z$ and $j''_z$ would vanish
and hence so would the resultant\footnote{By definition, the resultant 
of two polynomials 
\[
q_1 = a_n \Pi_{i=1}^n (x-\alpha_i),\qquad q_2 = b_m \Pi_{i=1}^m (x-\beta_i)
\]
is 
\[
R(q_1,q_2;x) = a_n^m b_m^n \Pi_{i=1}^n\Pi_{j=1}^m (\alpha_i - \beta_j)
\]
It can be computed from the Euclidean algorithm, or as the determinant of Sylvester's matrix
or Bezout's matrix.} of $q_1,q_2$, 
considering both as cubic polynomials
in~$x$.  
Up to an integer multiple, this resultant is computed\footnote{The necessary 
computations here and elsewhere in the paper are performed by the computer algebra package 
PARI using exact integer arithmetic.} to be
\[
	R(q_1,q_2;x) =
		y^{-34} (yz-1)^2
		p(y,z),
\]
where
\begin{equation}\label{def:p}
	\begin{array}{rcl}
	p(y,z) &=& \left(-1536\,{y}^{9}+4608\,{y}^{7} \right) {z}^{9}\\
&&{}+ \left( 3452\,{y}^{10}+2040\,{y}^{8}-20100\,{y}^{6} \right) {z}^{8}\\
&&{}+ \left( -3504\,{y}^{
11}-16456\,{y}^{9}-8608\,{y}^{7}+36120\,{y}^{5} \right) {z}^{7}\\
&&{}+
 \left( 1947\,{y}^{12}+20360\,{y}^{10}+62966\,{y}^{8}+48272\,{y}^{6}-
33769\,{y}^{4} \right) {z}^{6}\\
&&{}+ \left( -576\,{y}^{13}-11988\,{y}^{11}-
71800\,{y}^{9}-153832\,{y}^{7}\right.\\
&&\qquad -\left. 104440\,{y}^{5}+17900\,{y}^{3} \right) 
{z}^{5}\\
&&{}+ \left( 72\,{y}^{14}+3552\,{y}^{12}+38762\,{y}^{10}+143492\,{
y}^{8}\right.\\
&&\qquad+\left. 208760\,{y}^{6}+109100\,{y}^{4}-5530\,{y}^{2} \right) {z}^{4}\\
&&{}+
 \left( -432\,{y}^{13}+10464\,{y}^{11}-66316\,{y}^{9}-154672\,{y}^{7}\right. \\
&&\qquad-\left.
151552\,{y}^{5}-62848\,{y}^{3}+972\,y \right) {z}^{3}\\
&&{}+ \left( 1152\,{
y}^{12}+15384\,{y}^{10}+57803\,{y}^{8}+83120\,{y}^{6}\right. \\
&&\qquad+\left. 58958\,{y}^{4}+
20912\,{y}^{2}-81 \right) {z}^{2}\\
&&{}+ \left( -1440\,{y}^{11}-10536\,{y}^{9
}-21648\,{y}^{7}-20824\,{y}^{5}\right. \\
&&\qquad-\left. 11680\,{y}^{3}-3888\,y \right) z\\
&&{}+720\,
{y}^{10}+2160\,{y}^{8}+2760\,{y}^{6}+1908\,{y}^{4}+944\,{y}^{2}+324.
\end{array}
\end{equation}
Note that
\[
	q_1(x,1/z,z) = -8x^2(z^2-1)z^5\left(2xz^3-2xz+1\right),
\]
so it is zero only for $x=0$ or
$x = \frac{1}{2z-2z^3}$.
However, 
\[
q_2(1/(2z-2z^3),1/z,z)={\frac { \left( 5\,{z}^{2}-3 \right) {z}^{2}}{1-{z}^{2}}}.
\]
$z=0$ would correspond to $x=y=\infty$, which is not of interest, and
$\sqrt{3/5}>\half$.  It is thus enough to show that $p(y,z)$ has no real
zeros for $0\leq z \leq \half$.  This is easy to establish for $z=0$,
since
\begin{equation}\label{p0neg}
p(y,0) = 720\,{y}^{10}+2160\,{y}^{8}+2760\,{y}^{6}+1908\,{y}^{4}+944\,{y}^{2}+324 > 0
\end{equation}
for all $y$.
We postpone the rest of the proof of this proposition until the end of the section, and instead show first how this implies 
the desired result about $V$, i.e. that it satisfies condition (iii) of Theorem~\ref{thm:main}.
\begin{lemma} The function $j_0$  has no critical points other than at $y=0$. \end{lemma}

{\em Proof:}  We have for $z=0$
 \begin{equation}\label{atan}
\phi_0^{-1}(y) = y - \tan^{-1} y
\end{equation}
and
\[
q_1(x,y,0) = -4 y^{-10}x[(10y^4+35y^2+27)x^2-4(3y^3+2y^5)x+y^6].
\]
Consider the algebraic curve $q_1(x,y,0) = 0$.  It has a branch $x=0$. 
The other two branches can be transformed into a hyperbola by changing variables to 
\[
\xi := \frac{x}{y^3},\qquad \eta := \frac{x}{y}.
\]
Let $\mathfrak{H}$ denote the piece of this hyperbola that lies in the first quadrant of the 
$(\xi,\eta)$ plane, i.e. 
$${\mathfrak H} = \{(\xi,\eta)\ |\ \xi\geq 0, \eta\geq 0, h(\xi,\eta) = 0\},$$
 with
\begin{equation}\label{hyper}
h := 10\eta^2+35 \xi\eta + 27 \xi^2 - 12\xi-8\eta +1.
\end{equation}
Likewise, let $\mathfrak{T}$ denote the part of the transcendental curve $x = \phi_0^{-1}(y)$ 
 in the same quadrant:
$${\mathfrak T}=\{(\xi,\eta)\ |\ \xi\geq 0, \eta\geq 0, \tau(\xi,\eta) = 0\},$$
where $\tau$ has the following expansion
\[
\tau = -1 + \frac{1}{3\xi} - \frac{1}{5\xi^2}\eta + \frac{1}{7\xi^3}\eta^2 -\dots
\]
Recall that we have already shown by (\ref{p0neg}) that the function $j_0$ cannot have a stationary point of inflection.
Therefore on each branch of the hyperbola $\mathfrak H$ the second derivative $j_0''$ must be of one 
sign.  Thus one branch must correspond to minima and the other to maxima of $j_0$.

It is not hard to see that $\mathfrak{H}_L$, the lower branch  of the hyperbola $\mathfrak{H}$, is well separated from ${\mathfrak T}$: Let 
$R$ be the rectangle $[0,1/9]\times[0,6/37]$ in the $(\xi,\eta)$ plane.  Then 
\begin{claim} $\mathfrak{H}_L$ is contained in $R$ while $\mathfrak{T}\cap R = \emptyset$. \end{claim}
{\em Proof:}  The $\xi$-intercept of $\mathfrak{H}_L$ is at $\xi=1/9$.  The tangent line to $\mathfrak{H}_L$ at the point $(1/9,0)$ is $\eta = -\frac{54}{37}(\xi-\frac{1}{9})$.  Since $\mathfrak{H}_L$ is concave down, it lies below this tangent, and the $\eta$-intercept of the tangent line is at $\eta = 6/37$ which shows the first part of the claim.  Consider next the ODE satisfied by the curve 
$x = y - \tan^{-1} y$, i.e. $\frac{dx}{dy} = \frac{y^2}{y^2+1}$.  For $y\leq 1$ we have
$\frac{dx}{dy} \geq \frac{y^2}{2}$ which upon integration yields 
\begin{equation}\label{xibnd}
x(y) \geq \frac{y^3}{6},
\end{equation}
and thus $\xi \geq \frac{1}{6}$,
while if $y \geq 1$ then $\frac{dx}{dy} \geq \frac{1}{2}$ and thus, using the bound (\ref{xibnd}) at $y=1$,
\[
x(y) \geq \frac{1}{2}(y - 1) + x(1) \geq \frac{1}{2}(\frac{1}{3}y + \frac{2}{3}y) -
 \frac{1}{3} \geq \frac{1}{6}y,
\]
and thus $\eta \geq \frac{1}{6}$.  Thus along $\mathfrak{T}$ we have $\min(\xi,\eta)\geq 1/6$, which proves the second part of the claim.

Let $\mathfrak{H}_U$ denote the  upper branch of the hyperbola $\mathfrak{H}$.  It intersects $\mathfrak T$ at $\xi = \frac{1}{3}, \eta = 0$.  This point corresponds
to the origin of the $(x,y)$ plane where we know that $j_z$ has a local maximum. We have thus 
shown in the above Claim that $j_0'$ cannot have any local minima.  Moreover, we can compute the tangent lines
to the two curves at $(1/3,0)$ and we obtain that
\[
\left(\frac{d\eta}{d\xi}\right)_{\mathfrak{H}_U} = -\frac{18}{11},\qquad
\left(\frac{d\eta}{d\xi}\right)_{\mathfrak T} = -\frac{5}{3}.
\]
Thus $\mathfrak{H}_U$ is below $\mathfrak T$ near $(\frac{1}{3},0)$.  It is easy to see that this is also 
the case near the $\eta$ axis (which corresponds to the infinity of the $(x,y)$ plane).  Thus if
$\mathfrak T$ were to intersect $\mathfrak{H}_U$ it would have to do so at least twice. Since a continuous 
function cannot have two consecutive local maxima without a minimum in between 
this is not possible and the curves $\mathfrak{H}_U$ and $\mathfrak T$ have no
other common point in (the 
first quadrant of) the $(\xi,\eta)$ plane. This proves the lemma.
\qed

The following Proposition completes the proof of Theorem~\ref{thm:intro} by showing that  the quantity on the left in condition (iii) of Theorem~\ref{thm:main} is equal to $\frac{2}{9}$.

\begin{prop}\label{prop:jz}  The function $j_z$  for $0\leq z\leq 1/2$ has no local maximum other than at $y=0$. \end{prop}
{\em Proof:}
Let
\[
K := \{(y,z)| y > 0,\ j'_z(y) = 0,\ j''_z(y) \leq 0\}.
\]
Note that $K$ is closed.
From (\ref{taylor2}) we see that there is an $\epsilon >0$ such that there
$j'_z \neq 0$ in the region $0 < y < \epsilon$, $0 \le z \le \frac 1 2$.
Similarly, from (\ref{taylor5}) we see that there is a $Y$ such that
$j'_z \neq 0$ in the region $Y < y < \infty$, $0 \le z \le \frac 1 2$.
$K$ is therefore compact, since it
is a closed subset of $[\epsilon,Y]\times[0,\half]$.

If $K$ is nonempty, then there exists a point $(\bar{y},\bar{z})$
in $K$ with smallest $z$, i.e.
$z\geq \bar{z}$ for all $(y,z)\in K$. Thus
$j'_{\bar{z}}(\bar{y}) = 0$ and $j''_{\bar{z}}(\bar{y}) \leq 0$.
By the previous lemma, $\bar{z} > 0$.  Suppose that
$j''_{\bar{z}}(\bar{y}) \ne 0$.  Then by the implicit function theorem,
we can solve
\[
q_1(\phi_z^{-1}(y),y,z) = 0
\]
near $(\bar{y},\bar{z})$, to get the curve $y = \vartheta(z)$.
Now for $z<\bar{z}$, $(\vartheta(z),z)\notin K$ so  we have
$j''_z(\vartheta(z))>0$.
Thus by continuity, $j''_{\bar{z}}(\bar{y}) \geq 0$,
which leads to a contradiction.  Thus we must have
$j''_{\bar{z}}(\bar{y}) = 0$.
But this is ruled out by Proposition~\ref{prop:nospi}.
Therefore $K$ must be empty.  This proves the statement of Proposition~\ref{prop:jz}.
\qed

We conclude with the remainder of the proof of Proposition~\ref{prop:nospi}.
 First we need some definitions:  For
\[
p(y,z) = \sum_{k,l} c_{k,l} y^k z^l
\]
a polynomial in two variables with real coefficients, let $C$ be the curve
\[
C := \{(y,z) \in \RR^2 \ |\ p(y,z) = 0\}
\]
considered as a subset of $\RR^2$ in the usual topology.  The {\em Newton Polygon} of $p$ is defined to be the convex hull of the set 
\[
N := \{(k,l) \in \ZZ^2\ |\ c_{k,l} \ne 0\}.
\]
If $E$ is an (oriented) edge of the Newton polygon with endpoints $(k'_E,l'_E)$ and $(k''_E,l''_E)$, then the numbers
$d_E$, $p_E$ and $q_E$ are defined by 
\[
d_E := \gcd(k''_E-k'_E,l''_E-l'_E),\quad p_E := \frac{k''_E-k'_E}{d_E}, \quad q_E := \frac{l''_E-l'_E}{d_E},
\]
and the {\em edge polynomial} $e_E \in \RR[t]$ by
\[
e_E(t) :=\sum_{i=0}^{d_E} c_{k'_E+ip_E,l'_E+iq_E}t^i.
\]
An edge is called {\em outer} if it maximizes some linear function $ak+bl$ on the Newton polygon, where at 
least one of $a$ or $b$ is positive.  

The following  compactness criteria for plane algebraic curves is proved in \cite{Sta03b}:
\begin{theorem} 
For the compactness of $C$ it suffices that $p$ is not divisible by $y$ or $z$, and the
 edge polynomials corresponding to outer edges have no real zeros.
\end{theorem}

\begin{figure}[h]
\begin{center}
\includegraphics[width=0.5\textwidth]{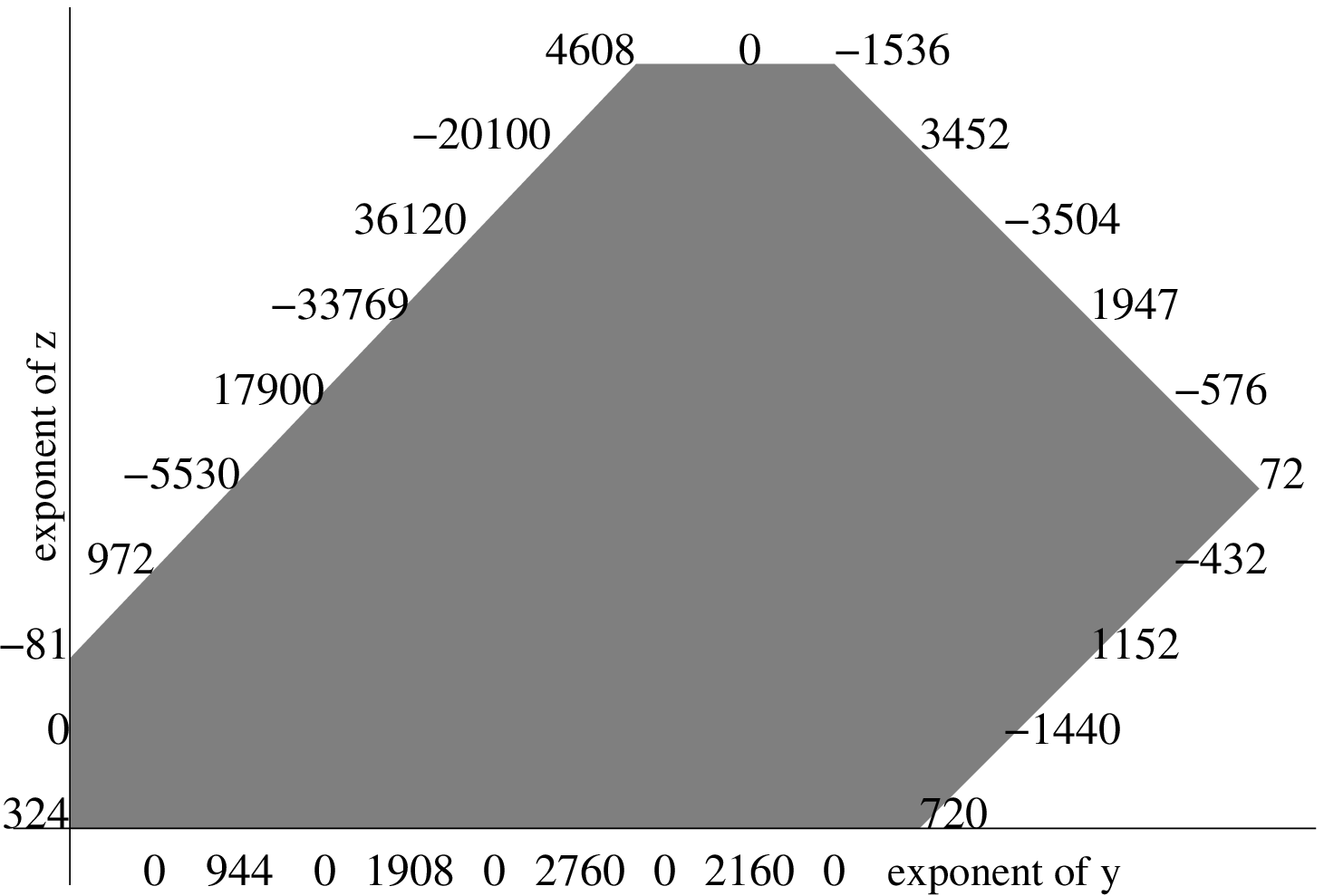}
\end{center}
\caption{\label{fig:np} Newton polygon of $p$}
\end{figure}

\begin{figure}[h]\begin{center}
\includegraphics[width=0.5\textwidth]{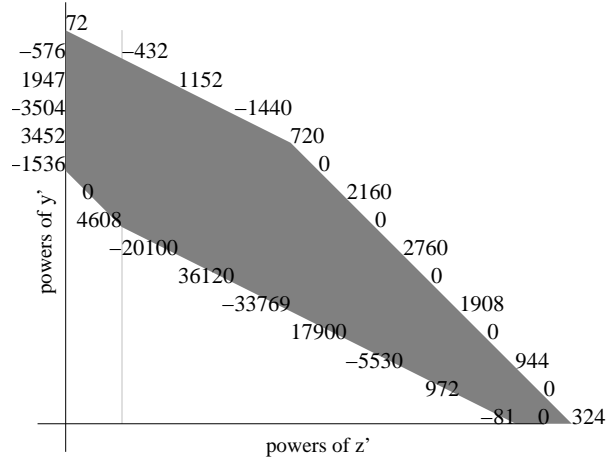}
\end{center}\caption{\label{fig:nq}  The Newton Polygon of $q$}
\end{figure}

The Newton polygon of the polynomial (\ref{def:p}) is shown in Figure~\ref{fig:np}.
It is  obvious that $p$ does not satisfy the criteria in this theorem, since some of 
the outer edge polynomials are of odd degree.  To remedy this, we take a simple projective 
transformation of the plane and consider its composition with $p$: Let
\begin{equation}\label{def:q}
q(z',y') := {z'}^{18}p(\frac{y'}{z'},\frac{1}{z'}).
\end{equation}
We need to show that $q$ has no zeros in the region $[2,\infty)\times(0,\infty)$.  
The Newton polygon of $q$ is the image of the Newton polygon of $p$ under the linear map $(k,l) \mapsto (18-k-l,k)$.  It is shown in Figure~\ref{fig:nq}.

There are two outer edges, with corresponding edge polynomials
\begin{eqnarray}
e_1(t) & = & 324 t^{10} + 994 t^8 + 1908 t^6 + 2760 t^4 + 2160 t^2 + 720,\\
e_2(t) & = & 72 t^4 - 432 t^3 + 1152 t^2 - 1440 t + 720.
\end{eqnarray}
$e_1$ is clearly never zero, and
the fact that $e_2$ has no real zeros is easily checked by Sturm's criterion.
  By the above Theorem,
 the zeroes of $q$ are  a compact set.  If
$z'$ is maximal for this set then $\partial q/\partial y'$ is zero.
The resultant of $q$ and $\partial q/\partial y'$ with respect to $y'$
must then be zero as well.
This resultant is
\[
	R(q,\frac{\partial q}{\partial y'};y') =
		z'^{114}({z'}-1)^{45}({z'}+1)^{45}(2{z'}-1)(2{z'}+1)f({z'})^3g({z'})
\]
where
\[
	\begin{array}{rcl}
	f({z'}) &=& 15268608{z'}^{12} - 91375200{z'}^{10} + 235796896{z'}^8
		- 336360313{z'}^6\\
&& {}
	 + 278925810{z'}^4	- 126777721{z'}^2 + 24542656
		\end{array}
\]

and
\[
	\begin{array}{rcl}
	 \lefteqn{g({z'})=}&&\\
		 2238642500162400000000{z'}^{32}
		&-&34444148848863120000000{z'}^{30}
		\cr 
		+ 237592851413120362800000{z'}^{28}
		&-& 985391370893335206960000{z'}^{26}
		\cr 
		+ 2763859141396512532788000{z'}^{24}
		&-& 5568163844214174878032000{z'}^{22}
		\cr 
		+ 8328390177670642537641736{z'}^{20}
		&-& 9407814473561334395291936{z'}^{18}
		\cr 
		+ 8074051365793494550609047{z'}^{16}
		&-& 5249226125431353947046641{z'}^{14}
		\cr 
		+ 2557646890465822492261299{z'}^{12}
		&-& 918262815118642547577717{z'}^{10}
		\cr 
		+ 238729228492213678314693{z'}^8
		&-& 44988866247608231183315{z'}^6
		\cr 
		+ 6473535589377087487753{z'}^4
		&-& 739775558130688228967{z'}^2 
		\cr 
		{}+ 49586421\lefteqn{501845352448.}&&
		\end{array}
\]
$f$ has no real zeros at all, while all the zeros of $g$
are contained in $|z'|<2$.  Both of these facts are established by Sturm's
criterion.  It follows that the maximal value of $z'$ is less that $2$, and this concludes the proof of
Proposition~\ref{prop:nospi}.
\qed

{\bf Acknowledgment}  This research was completed while the second author was a member at the Institute for Advanced Study,  Princeton NJ.  He wishes to thank the institute and its staff for their hospitality.

\bibliographystyle{plain}
\def\cprime{$'$}

\par\noindent $^{\mathbf a}$
%%%%%%%%%%%%%%%%%%%%%%% Adresse no 1 %%%%%%%%%%%%%%%%%
Department of Mathematics\\
Princeton University,\\
 Princeton NJ 08544\\
Email: {\sl stalker@math.princeton.edu}
\ \\
 \\ $^{\mathbf b}$
%%%%%%%%%%%%%%%%%%%%%%% Adresse no 2 %%%%%%%%%%%%%%%%%
Department of Mathematics\\
Rutgers, The State University of New Jersey\\
 110 Frelinghuysen Road, Piscataway NJ 08854\\
Email: {\sl shadi@math.rutgers.edu}

\end{document}